\documentclass[11pt,showpacs,showkeys,eqsecnum,footinbib,preprint,superscriptaddress]{revtex4}
\usepackage[utf8]{inputenc}
\usepackage[T1]{fontenc}
\usepackage{epsfig}
\usepackage{amsmath}
\usepackage{amsfonts}
\usepackage{amssymb}
\usepackage{color}
\usepackage{graphicx}
\usepackage{url,hyperref}
\usepackage{latexsym}
\usepackage{longtable}
\usepackage{float}
\usepackage{url}
\begin{document}
\title{\textsc{\Large {Closed analytical solutions of the d-dimensional Schr\"{o}dinger equation with deformed  Woods-Saxon potential plus double ring-shaped potential}}}
\author{ M.~Chabab}
\email{mchabab@uca.ma (Corresponding author)}
\affiliation{High Energy Physics and Astrophysics Laboratory, Department of Physics, Faculty of Sciences Semlalia, Cadi Ayyad University, P.O.B. 2390, Marrakesh 40000, Morocco.}
\author{A.~El Batoul}
\email{elbatoul.abdelwahed@edu.uca.ma}
\affiliation{High Energy Physics and Astrophysics Laboratory, Department of Physics, Faculty of Sciences Semlalia, Cadi Ayyad University, P.O.B. 2390, Marrakesh 40000, Morocco.}
\author{M.~Oulne}
\email{oulne@uca.ma}
\affiliation{High Energy Physics and Astrophysics Laboratory, Department of Physics, Faculty of Sciences Semlalia, Cadi Ayyad University, P.O.B. 2390, Marrakesh 40000, Morocco.}
\begin{abstract}
	\noindent
	By employing  the Pekeris approximation, the D-dimensional Schr\"{o}dinger equation is solved for the nuclear deformed  Woods-Saxon potential plus double ring-shaped potential within the framework of the Asymptotic Iteration Method (AIM). The energy eingenvalues are given in a closed form and the corresponding normalized eigenfunctions are obtained in terms of hypergeometric functions. Our general results  reproduce many predictions obtained in the literature, using the Nikiforov-Uvarov method (NU) and the Improved Quantization Rule approach, particulary those derived by considering Woods-Saxon potential without deformation and/or without ring shape interaction. 
\end{abstract}
\maketitle
\section{Introduction}
The Schr\"{o}dinger equation (SE) is one of the fundamental equations in quantum physics which still attracts strong interest of both physicists and mathematicians. However, solving this equation is often a complicated problem and exact analytic solutions may be found only in limited cases. Many advanced mathematical methods have been used to solve it. Among the most popular methods, one can cite the variational method \cite{b1}, the functional analysis method \cite{b2}, the supersymetric method \cite{b3}, the factorization method  \cite{b4}, the path integral method \cite{b5,b6,b7,b8}, the shifted 1/N expansion\cite{b9, b10}, the Nikiforov-Uvarov method (NU)\cite{b11,b12} and the quantization rule approach \cite{b13,b14}. Recently, the asymptotic iteration method (AIM) \cite{b15, b16, b17} an elegant, efficient technique to solve second-order homogeneous differential equations, has been the subject of extensive investigation in recent years, particularly when dealing withe non central potential. The Schr\"{o}dinger equation has been investigated for several potentials as the Woods-Saxon potential \cite{b18,b19,b20}, harmonic oscillator potential\cite{b21}, Hulthén potential \cite{b22,b23,b24,b25}, Kratzer potential \cite{b26}, generalized q-deformed Morse potential\cite{b27}, modifed Woods-Saxon potential\cite{b28}, Makarov potential \cite{b29}, deformed Woods-Saxon Potential\cite{b30}, Pseudoharmonic potential \cite{b31, b32}, Yukawa potential \cite{b33, b34} and Eckart potential \cite{b35,b36}.
Very recently, the Schr\"{o}dinger equation in generalized D dimensions for different potentials is getting more attention with the aim of generalizing the solutions to multidimensional space for many potentials \cite{b37,b38,b39,b40,b41,b42,b43,b44,b45,b46}. \\
In this paper we are interested in the Woods-Saxon potential \cite{b47}. This is one of the most realistic short-range potential in nuclear physics describing the central part of the interaction of single nucleon with whole nuclei. It is widely used in the study of nuclear structure within the shell model to obtain the nuclear energy level spacing, particle number dependence of energy quantities and universal properties of electron distributions in atoms, nuclei and atomic clusters \cite{b48,b49,b50} and also reproducing the magic numbers observed experimentally. Moreover, different generalized versions of this potential have been introduced to explore elastic and quasi-elastic scattering of nuclear particles \cite{b51}. On the other hand, the Woods-Saxon potential, either in its spherical or deformed form, has been used more in nuclear calculations \cite{b52,b53,b54,b55} , studying the behavior of valence electrons in metallic systems or in helium model \cite{b56} and in nonlinear scalar theory of mesons \cite{b57}. Very recently, it has been applied   appropriately to study the nuclear collective motion in the framework of the Bohr Hamiltonian \cite{b58}.
Besides, considerable efforts have been made to obtain  analytical solutions of the Schr\"{o}dinger equation or the Klein-Gordon equation with ring-shaped potentials \cite{b2,b59,b60,b61,b62,b63,b64,b65}. 
In particular a new ring-shaped potential has been introduced paving the way to new calculations in which this potential has been combined with the Coulomb potential \cite{b59}, Hulthén potential \cite{b60} and Kratzer potential \cite{b61}. Later on, a more general form of this ring shaped potential has been applied with a q-deformed Woods-Saxon potential \cite{b62}. Also,  Carpio-Bernido and  Bernido \cite{b63} have explored the Schr\"{o}dinger equation with a double ring-shaped oscillator via the path integral method. The same work has been conidered by Chang-Yuan et al. \cite{b2} in the framework of the functional analysis method. 
Very recently, we obtained general analytical solutions of Schr\"{o}dinger equation in the context of position dependent effective mass for a class of non central physical potentials \cite{b66}, involving in particular the Coulomb potential combined with several generalized versions forms of ring-shape such as double ring-shaped potential and novel angle-dependent (NAD) potential.
The results of these calculations with non-central potentials may have interesting applications, particularly in quantum chemistry such as the study of ring-shaped molecules like cyclic polyenes and benzene \cite{b61}.\\
In nuclear physics, the shape form of the potential also plays an important role particularly when studying the structure of deformed nuclei or the interaction between them. Therefore, our aim in the present work, is to investigate analytical bound state solutions of the  D-dimentional Schr\"{o}dinger equation with a general deformed Woods-Saxon potential combined to double ring-shaped potential reported in \cite{b2,b62,b66}, using the asymptotic iteration method.\\
The organization of the paper is as follows. In section II, we discuss the theoretical  background of the D-dimentional Schr\"{o}dinger equation. In section III, we derive the solutions of radial equation with a general deformed Woods-Saxon potential in multi dimensional space. In section IV, we solve the D-dimensional angular equations with double ring-shaped potential. The section V contains our discussion of some special cases and finally our conclusions is drawn in section VI.
\section{Theoretical background of the D-dimentional  Schr\"{o}dinger equation }
In spherical coordinates, the D-dimentional Schr\"{o}dinger equation for a particle in general non-central potential $V(r,\theta)$ can be written as \cite{b37,b38,b39}
\begin{align}
	&\left[\nabla_D^2+\frac{2M}{\hbar^2}\left(E-V(r,\theta)\right)\right]\Psi_{\ell_1 \dots \ell_{D-2}}^{(\ell_{D-1}=\ell)}\left(x\right)=0,\nonumber \\
	&\nabla_D^2=\frac{\partial^2}{\partial r^2}+\frac{\left(D-1\right)}{r}\frac{\partial}{\partial r}+\frac{1}{r^2}\left[ \frac{1}{\sin^{D-2}\left(\theta_{D-1}\right)}\frac{\partial}{\partial \theta_{D-1}}\left(\sin^{D-2}\left(\theta_{D-1}\right)\right)-\frac{L_{D-2}^2}{\sin^2\left(\theta_{D-1}\right)}\right]
	\label{E1}
\end{align}
where $E$ is the energy and $M$ is the reduced mass, $\hbar$ is the reduced Planck's constant and
$x$ is a D-dimensional position vector with the hyperspherical cartesian components
$x_1, x_2,\dots,x_D$. The relations between these cartesian coordinates and the hyperspherical coordinates $r$ and $\theta_j$ in D-dimensional space are defined by \cite{b37,b38,b39,b40,b41,b42}:
\begin{align}
	& x_1=r\cdot \cos\left(\theta_1\right)\sin\left(\theta_2\right)\dots \sin\left(\theta_{D-1}\right), \nonumber \\
	& x_2=r\cdot \sin\left(\theta_1\right)\sin\left(\theta_2\right)\dots \sin\left(\theta_{D-1}\right), \nonumber \\
	& x_3=r\cdot \cos\left(\theta_2\right)\sin\left(\theta_3\right)\dots \sin\left(\theta_{D-1}\right), \nonumber \\
	&\vdots \nonumber \\
	& x_j=r\cdot \cos\left(\theta_{j-1}\right)\sin\left(\theta_j\right)\dots \sin\left(\theta_{D-1}\right),\ 3\preceq j\preceq D-1, \nonumber \\
	&\vdots \nonumber \\
	& x_{D-1}=r\cdot \cos\left(\theta_{D-2}\right) \sin\left(\theta_{D-1}\right),\nonumber \\
	& x_{D}=r\cdot \cos\left(\theta_{D-1}\right),\ \sum_{j=1}^Dx_j^2=r^2,D\succeq 2.
	\label{E2}
\end{align}
For $D = 2$ we have $x_1=r\cos(\varphi) $, $x_2=r\sin(\varphi) $ and $x_1=r\cos(\varphi)\sin(\theta) $, $x_2=r\sin(\varphi)\sin(\theta)$ for $D=3$.
The Laplace operator $\nabla_D^2$ is given by \cite{b67}:
\begin{equation}
	\nabla_D^2= \sum_{j=1}^D\frac{\partial^2}{\partial x_j^2}\cdot
	\label{E3}
\end{equation}
The volume element of the configuration space is calculated as:
\begin{equation}
	dV_D=\prod_{j=1}^Dx_j=r^{D-1}drd\Omega,\ d\Omega=\prod_{j=1}^{D-1}\left(\sin\left(\theta_j\right)\right)^{j-1}d\theta_j,
	\label{E4}
\end{equation}
where $r\in [0,\infty[$, $\theta_1\in [0, 2\pi]$ and $\theta_j \in [0,\pi]$, $j\in [2,D-1]$. \\
In order to separate the variables in Eq. (\ref{E1}), and for a given angular momentum $\ell$, the total wave function $\Psi_{\ell_1 \dots \ell_{D-2}}^{(\ell)}(x)$ can be now decomposed as a product of the radial function $R_{\ell}(r)$ and the generalized spherical harmonics $Y_{\ell_1 \dots \ell_{D-2}}^{(\ell)}(\hat{x})$
\begin{equation}
	\Psi_{\ell_1 \dots \ell_{D-2}}^{(\ell)}(x)=R_{\ell}\left(r\right)Y_{\ell_1 \dots \ell_{D-2}}^{(\ell)}\left(\hat{x}\right),\ R_{\ell}\left(r\right)=r^{-\left(D-1\right)/2}U\left(r\right)
	\label{E5}
\end{equation}
In addition, the Eq. (\ref{E1}) permits a solution via separation of variables, if one writes generalized spherical harmonics $Y_{\ell_1 \dots \ell_{D-2}}^{(\ell)}\left(\hat{x}\right)$ \cite{b37,b38,b39,b40}:
\begin{align}
	& Y_{\ell_1 \dots \ell_{D-2}}^{(\ell)}\left(\hat{x}\right)=Y\left(\ell_1,\ell_2,\dots ,\ell_{D-2},\ell \right),\ \ell=|m| \ \ \rm{for} \ D=2, \nonumber \\
	& Y_{\ell_1 \dots \ell_{D-2}}^{(\ell)}\left(\hat{x}=\theta_1,\theta_2,\dots,\theta_{D-1}\right)=\prod_{j=1}^{D-1}H\left(\theta_j\right),
	\label{E6}
\end{align}
as simultaneous eigenfunctions of $L_j^2$ :
\begin{align}
	& L_1^2Y_{\ell_1 \dots \ell_{D-2}}^{(\ell)}\left(\hat{x}\right)=m^2Y_{\ell_1 \dots \ell_{D-2}}^{(\ell)}\left(\hat{x}\right), \nonumber \\
	& L_j^2Y_{\ell_1 \dots \ell_{D-2}}^{(\ell)}\left(\hat{x}\right)=\ell_j\left(\ell_j+j-1\right)Y_{\ell_1 \dots \ell_{D-2}}^{(\ell)}\left(\hat{x}\right),\ j\in [1,D-1],\nonumber \\ &\ell=0,1,\dots,\ell_{k}=0,1,\dots,\ell_{k+1}, \ k\in [2,D-2], \nonumber \\
	& \ell_1=-\ell_2,-\ell_2+1,\dots,\ell_2-1,\ell_2, \nonumber \\
	&  L_{D-1}^2Y_{\ell_1 \dots \ell_{D-2}}^{(\ell)}\left(\hat{x}\right)=\ell\left(\ell+D-2\right)Y_{\ell_1 \dots \ell_{D-2}}^{(\ell)}\left(\hat{x}\right). 
	\label{E7}
\end{align}
The unit vector along $x$ is usually denoted by $\hat{x}=x/r$. Additionally,
the angular momentum operators $L_j^2$ are defined as :
\begin{align}
	& L_1^2=-\frac{\partial^2}{\partial\theta_1^2},\nonumber \\
	& L_k^2=\sum_{a\prec b=2}^{k+1}L_{ab}^2=-\frac{1}{\sin^{k-1}\left(\theta_k\right)}\frac{\partial}{\partial\theta_k}\left(\sin^{k-1}\left(\theta_k\right)\frac{\partial}{\partial\theta_k}\right)+\frac{L_{k-1}^2}{\sin^2\left(\theta_k\right)},\ 2\preceq k\preceq D-1, \nonumber \\
	& L_{ab}=-i\left[x_a\frac{\partial}{\partial x_b}-x_b\frac{\partial}{\partial x_a}\right].
	\label{E8}
\end{align}
Let us turn to our study using a physical potential. The interaction in the nucleus is often described by using a potential which consists of Coulomb plus nuclear potentials.The latter is usually taken in the form of the Woods–Saxon type potential. Therefore the general deformed Woods-Saxon used in this work is obtained from the modified Woods-Saxon potential established in Ref \cite{b28} which is specified by two parameters $q_1$ and $q_2$. Thus, the general deformed Woods–Saxon potential plus double ring-shaped potential one is expressed as
\begin{align}
	 V(r,\theta)&=-\frac{V_0}{q_1+q_2e^{\frac{(r-R_0)}{a}}}+\frac{K}{r}+\frac{\hbar^2}{2M}\left(\frac{b}{r^2\sin^2(\theta)}+\frac{c}{r^2\cos^2(\theta)}\right), R_0=r_0A_0^{1/3},\nonumber \\ & R_0>>a, \ b\geq0,\ c\geq 0 \nonumber \\
	&=V_1(r)+\frac{\hbar^2}{2Mr^2}V_2(\theta)
	\label{E9}
\end{align}
where the first term in the right-hand side of Eq. (\ref{E9}) is a general deformed  Woods-Saxon (dWS), where $V_0$ is the depth of the potential, $q_1$ and $q_2$ are two real parameters which determine the shape (deformation) of the
potential, $a$ is the diffuseness of the nuclear surface, $R_0$ is the width of the potential, $A_0$ is the atomic mass number of the nucleus and $r_0$ is the radius parameter. The second term introduces the double ring-shaped potential.\\
The substitution of Eqs. (\ref{E3}) and  (\ref{E5})- (\ref{E9}) into Eq. (\ref{E1}) allows us to obtain, via the standard procedure of separation of variables, the following equation:
\begin{align}
	\bigg\{\frac{\partial^2}{\partial r^2}+\frac{\left(D-1\right)}{r}\frac{\partial}{\partial r}+\frac{1}{r^2}\left[ \frac{1}{\sin^{D-2}\left(\theta_{D-1}\right)}\frac{\partial}{\partial \theta_{D-1}}\left(\sin^{D-2}\left(\theta_{D-1}\right)\right)-\frac{L_{D-2}^2+b}{\sin^2\left(\theta_{D-1}\right)}-\frac{c}{\cos^2\left(\theta_{D-1}\right)}\right] \nonumber \\ +\frac{2M}{\hbar^2}\left(E+\frac{V_0}{q_1+q_2e^{\frac{\left(r-R_0\right)}{a}}}-\frac{K}{r}\right)\bigg\}  r^{-\left(D-1\right)/2}U\left(r\right)Y_{\ell_1 \dots \ell_{D-2}}^{(\ell)}\left(\hat{x}\right)=0, 
	\label{E10}
\end{align}
\begin{align}
	& L_{D-1}^2=-\frac{1}{\sin^{D-2}\left(\theta_{D-1}\right)}\frac{\partial}{\partial \theta_{D-1}}\left(\sin^{D-2}\left(\theta_{D-1}\right)\frac{\partial}{\partial \theta_{D-1}}\right)+\frac{L_{D-2}^2+b}{\sin^2\left(\theta_{D-1}\right)}+\frac{c}{\cos^2\left(\theta_{D-1}\right)}, \nonumber\\
	& L_{D-2}^2=-\frac{1}{\sin^{D-3}\left(\theta_{D-2}\right)}\frac{\partial}{\partial \theta_{D-2}}\left(\sin^{D-3}\left(\theta_{D-2}\right)\frac{\partial}{\partial \theta_{D-2}}\right)+\frac{L_{D-3}^2}{\sin^2\left(\theta_{D-1}\right)},\nonumber\\
	&\vdots \nonumber \\
	& L_{j}^2=-\frac{1}{\sin^{j-1}\left(\theta_{j}\right)}\frac{\partial}{\partial \theta_{j}}\left(\sin^{j-1}\left(\theta_{j}\right)\frac{\partial}{\partial \theta_{j}}\right)+\frac{L_{j}^2}{\sin^2\left(\theta_{j}\right)},\ j\in [2,D-2],\nonumber\\
	&\vdots \nonumber \\
	&	L_{1}^2=-\frac{\partial^2}{\partial\theta_1^2}.
	\label{E11}
\end{align}
where $L_k^2$, $k\in[1,D-1]$ are the angular operators. Thus, the last wave equation can be separated into the following radial and angular parts as:
\begin{align}
	&\left[\frac{d^2}{dr^2}+\frac{2M}{\hbar^2}\left(E+\frac{V_0}{q_1+q_2e^{\frac{\left(r-R_0\right)}{a}}}-\frac{K}{r}\right)-\frac{\tau}{r^2}\right]U\left(r\right)=0,\ \tau=\frac{\left(D+2\ell-1\right)\left(D+2\ell-3\right)}{4}
	\label{E12} 
\end{align}
\begin{align}
	\Bigg[\frac{1}{\sin^{D-2}\left(\theta_{D-1}\right)}\frac{d}{d \theta_{D-1}}\left(\sin^{D-2}\left(\theta_{D-1}\right)\frac{d}{d \theta_{D-1}}\right)+\ell\left(\ell+D-2\right)-\frac{\Lambda_{D-2}+b}{\sin^2\left(\theta_{D-1}\right)}
    \nonumber \\
	-\frac{c}{\cos^2\left(\theta_{D-1}\right)}\Bigg]H\left(\theta_{D-1}\right)=0,
	\label{E13} 
\end{align}
\begin{align}
	&\vdots \nonumber \\
	&\left[\frac{1}{\sin^{j-1}\left(\theta_{j}\right)}\frac{d}{d \theta_{j}}\left(\sin^{j-1}\left(\theta_{j}\right)\frac{d}{d \theta_{j}}\right)+\Lambda_{j}-\frac{\Lambda_{j-1}}{\sin^2\left(\theta_{j}\right)}\right]H\left(\theta_j\right)=0,\ j\in [2,D-2],
	\label{E14}\\
    &\vdots \nonumber \\
   &\left[\frac{d^2}{d\theta_1^2}+ \Lambda_1\right]H\left(\theta_1\right)=0,\ \Lambda_1=\ell_1^2=m^2.
	\label{E15}
\end{align}
where $\Lambda_p =\ell_p(\ell_p+p-1)$, $p\in[1,D-1]$ are separation constants. The
solution of (\ref{E15}) is periodic and must satisfy the periodic boundary condition  $H(\theta_1=\varphi) = H(\theta_1=\varphi+2\pi)$,  from which we obtain:
\begin{equation}
	H(\theta_1=\varphi)=\frac{1}{\sqrt{2\pi}}e^{\pm i\ell_1\varphi},\ \ell_1=0,1,2,\dots .
	\label{E16}
\end{equation}
Further, Eqs. (\ref{E13}) and (\ref{E14}) representing the angular wave equations become 
\begin{align}
	\frac{d^2H\left(\theta_{D-1}\right)}{d\theta_{D-1}^2}+\left(D-2\right)\frac{\cos\left(\theta_{D-1}\right)}{\sin\left(\theta_{D-1}\right)}\frac{dH\left(\theta_{D-1}\right)}{d\theta_{D-1}}+\Bigg[\ell\left(\ell+D-2\right)-\frac{\Lambda_{D-2}+b}{\sin^2\left(\theta_{D-1}\right)}
	\nonumber \\
	-\frac{c}{\cos^2\left(\theta_{D-1}\right)}\Bigg]H\left(\theta_{D-1}\right)=0,
	\label{E17}
\end{align}
\begin{equation}
	\frac{d^2H\left(\theta_{j}\right)}{d\theta_{j}^2}+\left(j-1\right)\frac{\cos\left(\theta_{j}\right)}{\sin\left(\theta_{j}\right)}\frac{dH\left(\theta_{j}\right)}{d\theta_{j}}+\left(\Lambda_j-\frac{\Lambda_{j-1}}{\sin^2\left(\theta_j\right)}\right)H\left(\theta_{j}\right)=0,\ j\in[2,D-2],\ D\succ 3.
	\label{E18}
\end{equation}
where $\Lambda_p$ is well-known parameter in three-dimensional space \cite{b67}.\\
Only Eqs.  (\ref{E12}) and (\ref{E17}-\ref{E18}) have to be solved. To this end, we use the asymptotic iteration method  and the first step consist in the conversion of these equations to standard forms suitable to AIM applications.
\section{The solutions of the radial equation}
To solve the radial equation (\ref{E12}), we introduce the following conversions 
\begin{equation}
	x=\frac{r-R_0}{R_0} ,\ \alpha=\frac{R_0}{a} 
	\label{E28}
\end{equation}
Thus the deformed Woods-Saxon potential in Eq (\ref{E9}) transforms into 
\begin{equation}
	V_{d-WS}=\frac{K}{R_0(1+x)}-\frac{V_0}{q_1+q_2e^{\alpha x}}
	\label{E29}
\end{equation}
Because Eq. (\ref{E12}) cannot be solved analytically due to the centrifugal term $\tau/r^2$ and the other term $K/r$ we have to use a proper approximation of these terms.
We now make use of the following expansions for the terms appearing in Eq. (\ref{E12}), namely
\begin{equation}
	V_1(r)=\frac{\tau}{r^2}=\frac{\tau}{R_0^2(1+x)^2}=\frac{\tau}{R_0^2}(1-2x+3x^2-4x^3+...)
	\label{E30}
\end{equation}
\begin{equation}
	V_2(r)=\frac{K}{R_0(1+x)}=\frac{K}{R_0}(1-x+x^2-x^3+...)
	\label{E31}
\end{equation}
Bearing in mind the Pekeris approximation \cite{b68,b69}, we shall replace the potential by $V_1^*$ and $V_2^*$:
\begin{equation}
	V_1^*(x)=\frac{\tau}{R_0^2}\left({c_0+\frac{c_1}{q_1+q_2e^{\alpha x}}+\frac{c_2}{(q_1+q_2e^{\alpha x})^2}}\right)
	\label{E32}
\end{equation}
\begin{equation}
	V_2^*(x)=\frac{K}{R_0}\left({d_0+\frac{d_1}{q_1+q_2e^{\alpha x}}+\frac{d_2}{(q_1+q_2e^{\alpha x})^2}}\right)
	\label{E33}
\end{equation}
After expanding Eqs. (\ref{E32}) and (\ref{E33}) in terms of $x^1$, $x^2$, $x^3$, $\cdots$ and next, comparing with Eqs. (\ref{E30}) and (\ref{E31}) respectively, we
obtain the expansion coefficients  $c_0$, $c_1$, $c_2$, $d_0$, $d_1$ and $d_2$ as follows:
\begin{align}
	& c_0=1+\left(\frac{q_1^2}{q_2^2}-\frac{2q_1}{q_2}-3\right)\alpha^{-1}+\left(\frac{3q_1^2}{q_2^2}+\frac{6q_1}{q_2}+3\right)\alpha^{-2}
	,\nonumber \\
	& c_1=\left(-\frac{2q_1^3}{q_2^2}+6q_1+4q_2\right)\alpha^{-1}+\left(-\frac{6q_1^3}{q_2^2}-\frac{18q_1^2}{q_2}-18q_1-6q_2\right)\alpha^{-2}
	, 
	\label{E34} \\
	& c_2=\left(\frac{q_1^4}{q_2^2}+\frac{2q_1^3}{q_2}-2q_1q_2-q_2^2\right)\alpha^{-1}+\left(\frac{3q_1^4}{q_2^2}+\frac{12q_1^3}{q_2}+18q_1^2+12q_1q_2+3q_2^2\right)\alpha^{-2}, \nonumber \\
	& d_0=1+\left(\frac{q_1^2}{2q_2^2}-\frac{q_1}{q_2}-\frac{3}{2}\right)\alpha^{-1}+\left(\frac{q_1^2}{q_2^2}+\frac{2q_1}{q_2}+1\right)\alpha^{-2}
	,\nonumber \\
	& d_1=\left(-\frac{q_1^3}{q_2^2}+3q_1+2q_2\right)\alpha^{-1}+\left(-\frac{2q_1^3}{q_2^2}-\frac{6q_1^2}{q_2}-6q_1-2q_2\right)\alpha^{-2}
	, 
	\label{E35}\\
	& d_2=\left(\frac{q_1^4}{2q_2^2}+\frac{q_1^3}{q_2}-q_1q_2-\frac{1}{2}q_2^2\right)\alpha^{-1}+\left(\frac{q_1^4}{q_2^2}+\frac{4q_1^3}{q_2}+6q_1^2+4q_1q_2+q_2^2\right)\alpha^{-2} \nonumber
\end{align}
Now, if we rewrite (\ref{E12}) by using a new variable of the form $z=\frac{1}{q_1+q_2e^{\alpha x}}$,  
we obtain
\begin{equation}
	\frac{d^2U(z)}{dz^2}+\frac{1-2q_1z}{z(1-q_1z)} \frac{dU(z)}{dz}+\frac{-\varepsilon^2+\beta^2z+\gamma^2z^2}{z^2(1-q_1z)^2}U(z)=0,
	\label{E36}
\end{equation}
where the following substitutions have been used,
\begin{equation}
	\varepsilon^2=\frac{2Ma^2}{\hbar^2}\left(-E+\frac{Kd_0}{R_0}\right)+\frac{\tau a^2c_0}{R_0^2},\ \beta^2=\frac{2Ma^2}{\hbar^2}\left(V_0-\frac{Kc_1}{R_0}\right)-\frac{\tau a^2d_1}{R_0^2} ,\  \gamma^2=\frac{2MV_0a^2d_2}{R_0\hbar^2}+\frac{\tau a^2c_2}{R_0^2}.
	\label{E37}
\end{equation}
with 
\begin{equation}
	\tau=\frac{\left(D+2\ell-1\right)\left(D+2\ell-3\right)}{4}
	\label{E38}
\end{equation}
The boundary conditions for the radial wave function are $U(0)=0$ and $ U(\infty)=0$ .\\
To solve the Eq.(\ref{E36})  by means of the asymptotic iteration method, we propose the following ansatz for the radial wave function,
\begin{equation}
	U(z)=z^{\mu}(1-q_1z)^{\sigma}f(z)
	\label{E39}
\end{equation}
with 
\begin{equation}
	\mu=\varepsilon ,\ \sigma=\sqrt{\varepsilon^2-\frac{\beta^2}{q_1}+\frac{\gamma^2}{q_1^2}}
	\label{E40}
\end{equation}
Hence, the equation (\ref{E36})  transforms into, 
\begin{equation}
	\frac{d^2f_n(z)}{dz^2}=\lambda_0(z)\frac{df_n(z)}{dz}+s_0(z)f_n(z)
	\label{E41}
\end{equation}
with
\begin{equation}
	\lambda_0(z)= \frac{2q_1\left(\mu+\sigma+1\right)-2\mu-1}{z(1-q_1z)} , \ s_0(z)=\frac{q_1\left(\mu+\sigma\right)\left(\mu+\sigma+1\right)-q_1\delta^2}{z(1-q_1z)},\ \delta^2=\frac{\gamma^2}{q_1^2}
	\label{E42}
\end{equation}
According to the AIM procedure, the energy eigenvalues are then computed by means of the termination condition (Eq. (2.13) of Ref.\cite{b15}). After a few iterations, the obtained solutions are 
\begin{align}
	& \mu_0=-\sigma_0-\frac{1}{2}\pm \frac{1}{2}\sqrt{1+4\cdot\delta^2},\nonumber \\ &\mu_1=-\sigma_1-\frac{3}{2}\pm  \frac{1}{2}\sqrt{1+4\cdot\delta^2},\\ &\mu_2=-\sigma_2-\frac{5}{2}\pm  \frac{1}{2}\sqrt{1+4\cdot\delta^2} ,\nonumber \\
	& \vdots \nonumber
	\label{E43}
\end{align}
from which we derive
\begin{equation}
	\mu_n=-\sigma_n-\left(n+\frac{1}{2}\right)\pm \frac{1}{2}\sqrt{1+4\cdot\delta^2},\ \ n=0,1,2,...
	\label{E44}
\end{equation}
Once the expressions of $\varepsilon$, $\beta$ and $\gamma$, given in Eq. (\ref{E37}), are substituted into Eq. (\ref{E44}), we get the generalized formula of the radial energy eigenvalues,
\begin{align}
	E_{n,\ell }^{(D)}=-\frac{V_0}{2q_1}+\frac{\hbar^2\left(D+2\ell-1\right)\left(D+2\ell-3\right)}{16MR_0^2}\cdot\frac{\left(2c_0q_1^2+c_1q_1+c_2\right)}{q_1^2}+\frac{K\left(2d_0q_1^2+d_1q_1+d_2\right)}{2R_0q_1^2}\nonumber \\ -\frac{\hbar^2}{8Ma^2}\cdot\left\lbrace N^2+\frac{\left\lbrace \frac{a^2\left(D+2\ell-1\right)\left(D+2\ell-3\right)\left(c_1q_1+c_2\right)}{4R_0^2q_1^2}-\frac{2Ma^2V_0}{\hbar^2q_1}+\frac{2MKa^2\left(d_1q_1+d_2\right)}{\hbar^2R_0q_1^2}\right\rbrace^2 }{N^2}\right\rbrace 
	\label{E45}
\end{align}
with
\begin{equation}
	N=\left(n+\frac{1}{2}\right)\pm\frac{1}{2}\sqrt{1+\frac{8MKa^2d_2}{\hbar^2R_0q_1}+\frac{a^2c_2\left(D+2\ell-1\right)\left(D+2\ell-3\right)}{R_0^2q_1}}
	\label{E46}
\end{equation}
The equation (\ref{E45}) can be written as
\begin{align}
E_{n,\ell }^{(D)}=&\frac{\hbar^2c_0\left(D+2\ell-1\right)\left(D+2\ell-3\right)}{8MR_0^2}+\frac{Kd_0}{R_0}-\frac{\hbar^2}{2Ma^2}\nonumber\\& \times\bigg\{\left(\frac{2n+1}{4}\right)\pm\frac{1}{4}\sqrt{1+\frac{8MKa^2d_2}{\hbar^2R_0q_1}+\frac{a^2c_2\left(D+2\ell-1\right)\left(D+2\ell-3\right)}{R_0^2q_1}}\nonumber\\&-\frac{\frac{a^2(c_1q_1+c_2)\left(D+2\ell-1\right)\left(D+2\ell-3\right)}{4R_0^2q_1^2}+\frac{2Ma^2(d_1q_1+d_2)K}{\hbar^2R_0q_1^2}-\frac{2Ma^2V_0}{\hbar^2q_1}}{\left(2n+1\right)\pm\sqrt{1+\frac{8MKa^2d_2}{\hbar^2R_0q_1}+\frac{a^2c_2\left(D+2\ell-1\right)\left(D+2\ell-3\right)}{R_0^2q_1}}}\bigg\}^2
\label{E47}
\end{align}
The corresponding  eigenfunctions (in the limit x$\rightarrow 0$) of equation (\ref{E41}) are the hypergeometrical functions,
\begin{equation}
	f(z)= {}_2F_1\left(-n,1+2\mu+2\sigma+n;2\mu+1;q_1z\right)
	\label{E48}
\end{equation}
where ${}_2F_1$ is a special case of the generalized hypergeometric function.
Therefore, according to the relation between hypergeometrical functions and the generalized Jacobi functions $P_n^{(\alpha,\beta)}(z)$  \cite{b70,b71}, the radial wave function can be written as
\begin{equation}
	U_{n\ell}(z)=A_{n\ell} \ z^{\varepsilon}\left(1-q_1z\right)^{\sqrt{\varepsilon^2-\frac{\beta^2}{q_1}+\frac{\gamma^2}{q_1^2}}}P_n^{\left(2\varepsilon,2\sqrt{\varepsilon^2-\frac{\beta^2}{q_1}+\frac{\gamma^2}{q_1^2}}\right)}\left(1-2q_1z\right)
	\label{E49}
\end{equation}
where  
\begin{equation}
	P_n^{(\alpha,\beta)}(1-2z)=\frac{1}{n!}z^{-\alpha}(1-z)^{\beta}\frac{d^n}{dz^n}[z^{n+\alpha}(1-z)^{n+\beta}]
	\label{E50}
\end{equation}
and $A_{n\ell}$ is a normalization constant to be determined from the normalization condition
\begin{equation}
	\int_{0}^{\infty}\left| R(r)\right|^2r^{D-1}dr=	\int_{0}^{\infty}\left| U(r)\right|^2dr=1 
	\label{E51}
\end{equation}  
If we introduce a new variable $s=q_1z$, then the normalization condition (\ref{E51}) reduces to
\begin{equation}
	A_{n\ell}^2\int_{0}^{p}s^{2\mu-1}(1-s)^{2\sigma-1}\left[_2F_1\left(-n,1+2\mu+2\sigma+n;2\mu+1;s\right)\right]^2ds=\frac{q_1^{2\mu}}{a}, 
	\label{E52}
\end{equation}
where $p=\frac{1}{1+\frac{q_2}{q_1}e^{-\frac{R_0}{a}}}\rightarrow 1$ $(R_0>>a)$.\\
Using the following series representation of the hypergeometric function
\begin{equation}
	{}_p{F}_q\left(a_1,\cdots,a_p;c_1,\cdots,c_q;z\right)=\sum_{n=0}^{\infty}\frac{(a_1)_n\cdots(a_p)_n}{(c_1)_n\cdots(c_q)_n}\frac{z^n}{n!}
	\label{E53}
\end{equation}
where the Pochhammer symbols $(a)_i$ are defined by  
\begin{equation}
	(a)_i=\frac{\Gamma(a+i)}{\Gamma(a)}
	\label{E54}
\end{equation}
so we arrived at 
\begin{equation}
	A_{n\ell}^2\sum_{i=0}^{n}\sum_{j=0}^{n}\frac{(-n)_i(2\mu+2\sigma+n+1)_i}{(2\mu+1)_ii!}\frac{(-n)_j(2\mu+2\sigma+n+1)_j}{(2\mu+1)_jj!}\int_{0}^{1}s^{2\mu+i+j-1}(1-s)^{2\sigma-1}ds=\frac{q_1^{2\mu}}{a}
	\label{E55}
\end{equation}
where $2\mu+i+j>0$ and $2\sigma>0$.\\
Hence, by the definition of the Beta function, the equation (\ref{E55}) becomes
\begin{equation}
	A_{n\ell}^2\sum_{i=0}^{n}\sum_{j=0}^{n}\frac{(-n)_i(2\mu+2\sigma+n+1)_i}{(2\mu+1)_ii!}\frac{(-n)_j(2\mu+2\sigma+n+1)_j}{(2\mu+1)_jj!}B(2\mu+i+j,2\sigma)=\frac{q_1^{2\mu}}{a}
	\label{E56}
\end{equation}
where 
\begin{equation}
	B(x,y)=\int_{0}^{1}s^{x-1}(1-s)^{y-1}ds=\frac{\Gamma(x)\Gamma(y)}{\Gamma(x+y)},\ Re(x),Re(y)>0.
	\label{E57}
\end{equation}
Thus, the normalization constant $A_{n\ell}$ is now obtained as 
\begin{equation}
	A_{n\ell}=\left[\frac{q_1^{2\mu}}{a\cdot Q_{n}^{(\mu,\sigma)}}\right]^{1/2}
	\label{E58}
\end{equation}
where
\begin{equation}
	Q_{n}^{(\mu,\sigma)}=\sum_{i=0}^{n}\frac{(-n)_i(2\mu+2\sigma+n+1)_i}{(2\mu+1)_ii!}\times\sum_{j=0}^{n}\frac{(-n)_j(2\mu+2\sigma+n+1)_j}{(2\mu+1)_jj!}\frac{\Gamma(2\mu+i+j)\Gamma(2\sigma)}{\Gamma(2\mu+2\sigma+i+j)}
	\label{E59}
\end{equation}
\section{The solutions of the D-dimensional angular	equations}
In order to apply the AIM, we introduce a new variable $y =\cos(\theta_j)$. As a consequence, Eq. (\ref{E18}) is rearranged in the form of the universal associated-Legendre differential equation:
\begin{equation}
	\frac{d^2H(y)}{dy^2}-\frac{jy}{1-y^2}\frac{dH(y)}{dy}+\frac{\Lambda_j-\Lambda_{j-1}-\Lambda_jy^2}{\left(1-y^2\right)^2}H(y)=0,\ j\in[2,D-2],\ D\succ 3. 
	\label{E60}
\end{equation}
Now, we use the following ansatz for the  angular wave functions, 
\begin{equation}
	H(y)=(1-y^2)^\eta \xi(y)
	\label{E61}
\end{equation}
with
\begin{equation}
	\eta=\frac{(2-j)+\sqrt{\left(j-2\right)^2+4\Lambda_{j-1}}}{4}
	\label{E62}
\end{equation}
The equation (\ref{E60})  transforms into, 
\begin{equation}
	\frac{d^2\xi_{n_j}(y)}{dy^2}=\lambda_0(y)\frac{d\xi_{n_j}(y)}{dy}+s_0(y)\xi_{n_j}(y)
	\label{E63}
\end{equation}
with
\begin{equation}
	\lambda_0(y)=\frac{\left(4\eta+j\right)y}{1-y^2},\  s_0(y)=\frac{4\eta^2+2\left(j-2\right)\eta-\Lambda_j}{1-y^2}
	\label{E64}
\end{equation}
According to the AIM procedure, the energy eigenvalues are then computed by means of the quantization condition (Eq. (2.13) of Ref.\cite{b15}). After few iterations, the obtained solutions are  
\begin{align}
	&\eta_0=-\frac{-1+j}{4}+\frac{\sqrt{\left(j-1\right)^2+4\Lambda_j}}{4},\nonumber \\  &\eta_1=-\frac{1+j}{4}+\frac{\sqrt{\left(j-1\right)^2+4\Lambda_j}}{4}, 
	\label{E65} \\ 
	&\eta_2=-\frac{3+j}{4}+\frac{\sqrt{\left(j-1\right)^2+4\Lambda_j}}{4},\nonumber \\  
	&\vdots \nonumber
\end{align}
In general form, we have
\begin{equation}
	\eta_{n_j}=-\frac{(2n_j-1)+j}{4}+\frac{\sqrt{\left(j-1\right)^2+4\Lambda_j}}{4},\ n_j=0,1,2,\cdots
	\label{E66}
\end{equation}
Substituting the expression of $\eta$ given by Eq. (\ref{E62}) we get:
\begin{equation}
	\frac{\sqrt{\left(j-1\right)^2+4\Lambda_j}}{2}-\frac{\sqrt{\left(j-2\right)^2+4\Lambda_{j-1}}}{2}-\frac{1}{2}=n_j,\ n_j=0,1,2,\cdots
	\label{E67}
\end{equation}
with $\Lambda_j =\ell_j(\ell_j+j-1)$. This equation is exactly similar to the one obtained in \cite{b40,b41,b42}. The corresponding  eigenfunctions  of equation (\ref{E63}) are the universal associated-Legendre polynomials:
\begin{equation}
	\xi_{n_j}(y)=\frac{(-1)^{\tilde{m}}}{2^{\tilde{n_{j}}}\tilde{n_{j}}!}\frac{d^{\tilde{n_{j}}+\tilde{m}}}{dy^{\bar{n_{j}}+\tilde{m}}}\left(y^2-1\right)^{\tilde{n_{j}}}
	\label{E68}
\end{equation}
where $\tilde{n_{j}}$ and $\tilde{m}$ are given by :
\begin{equation}
	\tilde{n_{j}}=n_{j}+\frac{(j-2)}{2}+2\eta,\ \tilde{m}=\frac{(j-2)}{2}+2\eta
	\label{E69}
\end{equation}
Finally the angular wavefunction is
\begin{equation}
	H(y)=\frac{(-1)^{\tilde{m}}}{2^{\tilde{n_{j}}}\tilde{n_{j}}!}\left(1-y^2\right)^{\eta}\frac{d^{\tilde{n_{j}}+\tilde{m}}}{dy^{\bar{n_{j}}+\tilde{m}}}\left(y^2-1\right)^{\tilde{n_{j}}},\ j\in[2,D-2],\ D\succ 3.
	\label{E70}
\end{equation}
where the $n_j$, given in Eq. (\ref{E67}), becomes
\begin{equation}
	n_j=\ell_j-\ell_{j-1}, \ j\in[2,D-2],\ D\succ 3.
	\label{E71}
\end{equation}
Similarly, to calculate the eigenvalues of the angular equation (\ref{E17}), we introduce a new variable $t=\cos(\theta_{D-1})$ so we obtain the following equation 
\begin{equation}
	\frac{d^2H(t)}{dt^2}-\frac{(D-1)t}{1-t^2}\frac{dH(t)}{dt}-\frac{\left(\ell\left(\ell+D-2\right)\right)t^4+\left(b-c+\Lambda_{D-2}-\ell\left(\ell+D-2\right)\right)t^2+c}{t^2\left(1-t^2\right)^2}H(t)=0,
	\label{E72}
\end{equation}
To solve this equation by means of the AIM, we propose the following ansatz for the polar angular wave functions, 
\begin{equation}
	H(t)=t^{\alpha}(1-t^2)^{\beta}\chi(t)
	\label{E73}
\end{equation}
where
\begin{equation}
	\alpha=\frac{1+\sqrt{1+4c}}{2}, \ \beta=\frac{(3-D)+\sqrt{(D-3)^2+4(b+\Lambda_{D-2})}}{4}
	\label{E74}
\end{equation}
The equation (\ref{E72}) reduces to, 
\begin{equation}
	\frac{d^2\chi_{n_{D-1}}(t)}{dt^2}=\lambda_0(t)\frac{d\chi_{n_{D-1}}(t)}{dt}+s_0(t)\chi_{n_{D-1}}(t)
	\label{E75}
\end{equation}
with
\begin{equation}
	\lambda_0(t)= \frac{\left(2\left(\alpha+2\beta\right)+D-1\right)t^2-2\alpha}{t(1-t^2)},\
	s_0(t)=\frac{\left(\alpha+2\beta-\ell\right)\left(\alpha+2\beta+\ell+D-2\right)}{1-t^2}
	\label{E76}
\end{equation}
Using the AIM procedure for this equation, the energy eigenvalues are then obtained. So, we have
\begin{equation}
	\ell_0=2\beta+\alpha,\ \ell_1=2+2\beta+\alpha,\ \ell_2=4+2\beta+\alpha,\ \cdots 
\end{equation}
from which we find the generalized  relation of quantum number $\ell$:
\begin{equation}
	\ell_{n_{D-1}}=2\beta+\alpha+2n_{D-1} ,\ n_{D-1}=0,1,2,\cdots
	\label{E78}
\end{equation}
Substituting the obtained expression for $\beta$ and $\alpha$ given in Eq.(\ref{E74}) into Eq. (\ref{E78}) we get
\begin{equation}
	\ell_{n_{D-1}}=\frac{(4-D)+\sqrt{1+4c}+\sqrt{(D-3)^2+4(b+\Lambda_{D-1})}}{2}+2n_{D-1}, \ b\neq0, c>0, n_{D-1}=0,1,2,\cdots
	\label{E79}
\end{equation}
which is the same as the result reported in \cite{b2,b66} for $D=3$.\\
The eigenfunctions of equation (\ref{E75}) are the hypergeometrical functions. 
\begin{equation}
	\chi(t)={}_2F_1\left(-n_{D-1},\alpha+2\beta+n_{D-1}+\frac{(D-2)}{2};\alpha+\frac{1}{2};t^2\right) 
	\label{E80}
\end{equation}
Finally we get the exact eigenfunctions  of equation(\ref{E72}) 
\begin{equation}
	H(t)=C_{n_{D-1}}t^{\alpha}\left(1-t^2\right)^{\beta}{}_2F_1\left(-n_{D-1},\alpha+2\beta+n_{D-1}+\frac{(D-2)}{2};\alpha+\frac{1}{2};t^2\right) 
	\label{E81}
\end{equation}
where $C_{n_{D-1}}$ is a normalization constant computed via the orthogonality relation of Jacobi polynomials \cite{b70,b71}
\begin{equation}
	C_{n_{D-1}}=\left[\frac{\left(2n_{D-1}+\alpha+2\beta+\frac{(D-2)}{2}\right)\Gamma\left( n_{D_1}+\alpha+2\beta+\frac{(D-2)}{2} \right)\Gamma\left(n_{D-1}+\alpha+\frac{1}{2}\right)}{n_{D-1}!\Gamma\left(n_{D-1}+2\beta+\frac{D-1}{2}\right)\Gamma\left(\alpha+\frac{1}{2}\right)^2}\right]^{1/2}
	\label{E82}
\end{equation}
where $\alpha$ and $\beta$ are given in Eq. (\ref{E74}).\\
Now, in order to solve the angular equation (\ref{E72}), in the special case where $c = 0$ and $b\geq0$ we take
\begin{equation}
	\alpha=0,\ \beta=\frac{(3-D)}{4}+\frac{\sqrt{(D-3)^2+4(b+\Lambda_{D-2})}}{4}
	\label{E83}
\end{equation}
Then the equation (\ref{E75}) can be written as
\begin{equation}
	\frac{d^2\chi_{n_{D-1}}(t)}{dt^2}=\lambda_0(t)\frac{d\chi_{n_{D-1}}(t)}{dt}+s_0(t)\chi_{n_{D-1}}(t)
	\label{E84}
\end{equation}
with
\begin{equation}
	\lambda_0(t)=\frac{(D-1+4\beta)t}{1-t^2} ,\
	s_0(t)=\frac{(2\beta-\ell)(2\beta+\ell+D-2)}{1-t^2}
	\label{E85}
\end{equation}
Following the same procedure as in the previous case, we have: 
\begin{equation}
	 \ell_0=2\beta,\ \ell_1=2\beta+1,\ \ell_2=2\beta+2,\ \ell_3=2\beta+3,\ \cdots 
\end{equation}
In general form, we get:
\begin{equation}
	\ell_{n_{D-1}}=2\beta+n_{D-1}, \ n_{D-1}=0,1,2,3,...
	\label{E87}
\end{equation}
After substituting the expression of $\beta$ in Eq. (\ref{E83}), we obtain:
\begin{equation}
	\ell_{n_{D-1}}=\frac{(3-D)}{2}+\frac{\sqrt{(D-3)^2+4(b+\Lambda_{D-2})}}{2}+n_{D-1}, \ c=0,\ b\geq0, n_{D-1}=0,1,2,...
	\label{E88}
\end{equation}
Additionally, the differential equation (\ref{E75}) reduces to a universal associated Legendre differential equation, whose solutions are given by universal associated Legendre polynomials \cite{b70,b71} :
\begin{equation}
	\chi_{n_{D-1}}(t)=\frac{(-1)^{\hat{m}}}{2^{\hat{\ell}}\hat{\ell}!}\frac{d^{\hat{\ell}+\hat{m}}}{dt^{\hat{\ell}+\hat{m}}}\left(t^2-1\right)^{\hat{\ell}}
	\label{E89}
\end{equation}
where $\hat{\ell}$ and $\hat{m}$ are given by :
\begin{equation}
	\hat{\ell}=\ell_{n_{D-1}}+\frac{(D-3)}{2},\ \hat{m}=\frac{D-3}{2}+2\beta
	\label{E90}
\end{equation}
Finally in this cas with  $c = 0$ and $b\geq0$, we obtain the exact eigenfunctions of equation (\ref{E72}) 
\begin{equation}
	H(t)=\frac{(-1)^{\hat{m}}}{2^{\hat{\ell}}\hat{\ell}!}\left(1-t^2\right)^{\beta}\frac{d^{\hat{\ell}+\hat{m}}}{dt^{\hat{\ell}+\hat{m}}}\left(t^2-1\right)^{\hat{\ell}}
	\label{E91}
\end{equation}
where $\beta$ is given in Eq. (\ref{E83}).
\section{Discussion}
In this section, we consider some special cases of the potential under consideration. So in the case where the double ring-shaped potential vanishes $(ie :c=b=0)$, the general expressions obtained, in this work, for the energy spectrum reproduce easily the results derived in many particular cases. Indeed,
\begin{itemize}
	\item (i): Setting $q_1=q_2=1$ and $K=0$ the potential in Eq.(\ref{E9}) reduces to 
	\begin{equation}
		V_1(r)=-\frac{V_0}{1+e^{\frac{(r-R_0)}{a}}}
		\label{E92}
	\end{equation} 
	which is known as the standard Woods-Saxon potential. In this case we obtain energy eigenvalues and eigenfunctions identical to those reported in the Ref. \cite{b72,b73} via Nikiforov-Uvarov Method(NU) for $D=3$.
	\item (ii): In the case of $q_1=q_2=1$ and $K\ne0$, the potential in Eq. (\ref{E9}) becomes:
	\begin{equation}
		V_1(r)=-\frac{V_0}{1+e^{\frac{(r-R_0)}{a}}}+\frac{K}{r}
		\label{E93}
	\end{equation} 
 known as modified Woods-Saxon potential. Also here our expressions for the energy eigenvalues and the corresponding eigenfunctions reproduce exactly the results found in \cite{b28} with Nikiforov-Uvarov Method (NU).
	\item (iii): Finally, if we set $K=0$, $q_1=q$ and $q_2=1$  the potential (\ref{E9}) reads as
	\begin{equation}
		V_1(r)=-\frac{V_0}{q+e^{\frac{(r-R_0)}{a}}}=-\frac{V_0e^{-\frac{(r-R_0)}{a}}}{1+qe^{-\frac{(r-R_0)}{a}}}
		\label{E94}
	\end{equation} 
	which is known as the deformed Woods-Saxon potential. Again, we obtain an energy sprectrum similar to the sprectrum  reported in \cite{b42} using Improved Quantization Rule.\\
	On the other hand, when we set the parameters  $c_0 = \alpha^2/12$, and
	$c_1 = c_2 = \alpha^2$ (using the same approximation invoked in \cite{b42} to deal with the centrifugal term $1/r^2$), and by fixing $q=-1$, $R_0 =1$ and $V_0 =\alpha Ze^2$  in Eq. (\ref{E47}), we recover the energy spectrum of the Hulthén potential. 
\end{itemize}
\section{Conclusion}
In this work, the bound states of the D-dimensional Schr\"{o}dinger equation for a general deformed Woods-Saxon potential plus double ring-shaped potential  are investigated. Using the asymptotic iteration method, we have derived the general expression of the energy eigenvalues and the corresponding normalized eigenfunctions in terms of the hypergeometrical functions in a multi-dimensional space. Also we have shown that our generalized results reproduce the energy spectrum in many particular cases obtained previously in literature by other methods. Finally, the obtained theoretical results of the present problem may find applications in several fields of physics, especially in quantum chemistry and in the study of the interaction between deformed nuclear particles. 


\end{document}